\documentclass[aps,prc,superscriptaddress,showpacs,nofootinbib]{revtex4}
\usepackage[dvips]{graphicx}
\usepackage{amsmath,amssymb}

\newcommand{\Btl}{\mbox{${\tilde N}_B$}}

\newcommand{\tlX}{\mbox{${\tilde X}$}}

\newcommand{\tlT}{\mbox{${\tilde T}$}}
\newcommand{\tlV}{\mbox{${\tilde V}$}}

\newcommand{\vx}{\mbox{\boldmath $x$}}

\newcommand{\vcr}{\mbox{\boldmath $r$}}

\newcommand{\vlambda}{\mbox{\boldmath $\lambda$}}
\newcommand{\tvlamb}{\mbox{\boldmath $^t\lambda$}}
\newcommand{\dvlamb}{\mbox{\boldmath ${\dot \lambda}$}}
\newcommand{\tdvlamb}{\mbox{\boldmath $^t{\dot \lambda}$}}

\begin{document}

\title{\bf
Coupled Breathing Oscillations of Two-Component Fermion Condensates \\
in Deformed Traps}

\author{Tomoyuki~Maruyama}
\affiliation{College of Bioresource Sciences,
Nihon University,
Fujisawa 252-8510, Japan}
\affiliation{Advanced Science Research Center,
Japan Atomic Energy Research Institute,Tokai 319-1195, Japan}
\affiliation{Department of Physics, Tokyo Metropolitan University, 
 Hachioji, Tokyo 192-0397, Japan}

\author{Takushi Nishimura}
\affiliation{Department of Physics, Tokyo Metropolitan University, 
 Hachioji, Tokyo 192-0397, Japan}

\begin{abstract}
We investigate collective excitations coupled with monopole and 
quadrupole oscillations in two-component fermion condensates in deformed traps.
The frequencies of monopole and dipole modes
are calculated using Thomas-Fermi theory and the scaling approximation.
When the trap is largely deformed, these collective motions are decoupled
to the transverse and longitudinal breathing oscillation modes.
As the trap approaches becoming spherical, however,
they are coupled and show complicated behaviors.
\end{abstract}


\maketitle

\section{Introduction}

Since the realization of the Bose-Einstein condensed (BEC) atomic 
gases \cite{BECrv,nobel}, 
there has been much interest in ultracold trapped atomic systems 
to study quantum many-body phenomena. 
Besides the Bose-Einstein condensates (BEC) \cite{BECrv,nobel,bec}, 
one can now study degenerate atomic Fermi gases \cite{ferG} 
and bose-fermi mixtures \cite{BferM}.
These systems offer great promise to exhibit new and interesting
phenomena of quantum many-particle physics.

Sogo and Yabu \cite{SY} studied the ground state properties of fermi gas composed by two
spin components with various repulsive interaction and 
showed the phase transition between the paramagnetic and ferromagnetic states.
Furthermore we extended this study on the spin phases of the ground state to that
for the asymmetric gas with a nets spin in the presence 
of an external magnetic field and gave the phase diagram about these
spin phases  \cite{ToBe}.

An important diagnostic signal for many-body systems 
is the spectrum of collective excitations.
Such oscillations are common to a variety of many-particle systems
and are often sensitive to the interaction and the structure of the
ground state.  
In ref.\cite{ToBe} we have also studied 
the spin excitation on the dipole and
monopole oscillations in the two-component fermion condensed system.
The frequencies of the collective oscillations with the out-of-phase
show the dependence on the fermion-fermion interaction while
those with the in-phase are not sensitive to the interaction. 
Particularly the frequencies of the out-of-phase modes
show rapid changes in the phase transition between the paramagnetic and
ferromagnetic spin-phases. 

The monopole oscillations can be defined only 
in the spherical symmetric system.
In usual experiments traps have elliptically deformed shapes, where
the monopole oscillations are coupled with
the quadrupole oscillations.
If the trap is largely deformed with the axial symmetry
the oscillations must be decoupled into the modes of oscillations along 
the transverse and longitudinal  directions;
we shall call them the  transverse breathing (TB) 
and longitudinal breathing (LB) modes.

In this paper we study the coupled oscillation states 
of the two component fermi gases between the TB and LB modes
in deformed traps.
The oscillation behaviors are expected to show various features  
as the deformation of the trap is changed.

The underlying theoretical tool to treat dynamic problems of
dilute quantum gases is the time-dependent mean field theory.  
This is reduced to the random-phase approximation (RPA) for small
amplitudes, and the theory in this form has been applied to
density oscillations of these systems \cite{ClFer,Kry,ClSp}.  
When the
single-particle spectrum is regular, the long-wavelength excitations 
are collective and simpler methods can be used to calculate the
frequencies, in particular with sum rules \cite{sumF} 
or the scaling approximation. 
In this work we calculate the frequencies of these collective oscillations 
by using the scaling method \cite{scal,bohi}.
This scaling method is physically quite
transparent, but it is in fact equivalent to 
the theory based on energy-weighted
sum rules \cite{bohi}, as used for example in ref.~\cite{sumF}.

We will only consider here the case of positive scattering lengths which 
correspond to a repulsive interaction.  
When the interaction is
attractive, the system is superfluid in its ground state and the
excitation properties are controlled by the energy gap.
In addition we will focus only on the collective oscillations 
in the paramagnetic spin phases, and do not deeply discuss those 
in the ferromagnetic regime.  

In the next section we will explain our formulation, 
and show calculational results in Sec. 3.
In Sec.4 we will summarize our work.

\section{Formalism}
\label{grdSec}

We begin with the expression for the energy of a dilute trapped 
condensate in the mean-field approximation:
\begin{eqnarray}
E_T &=& \int d^3 r ~ [~ \frac{\hbar^2}{2m} 
 \sum_{s=1,2} \sum_{j=1}^{3} \tau_{s,j}
+ \frac{m}{2} ( \Omega_T^2 r_1^2 + \Omega_T^2 r_2^2 + \Omega_L^2 r_3^2 )
(\rho_1+\rho_2)  + g_F \rho_1\rho_2  ~] 
\label{etot}
\end{eqnarray}
with
\begin{eqnarray}
\tau_{s,j} &=& \sum_n^{occ} 
\frac{\partial \psi^*_{n,s}}{\partial r_j} 
\frac{\partial \psi_{n,s}}{\partial r_j}
\\
\rho_s &=& \sum_n^{occ} \psi^*_{n,s} (\vcr) \psi_{n,s} (\vcr),
\end{eqnarray}
where $\psi_{n s}$ are orbital wave functions indexed by orbit $n$ 
and component $s=1,2$,
$\Omega_{L,T}$ are the longitudinal and transverse frequencies of the
trapping field and $g_F$ is the
coupling strength of a contact interaction.
Here we assume that all fermions have the same mass, $m$,
but we do not need to treat these components as the spin state of
the same fermion.

In order to study the collective oscillation,
we introduce the following scaling for the fermion wave-function:
\begin{equation}
\psi_{n,s}(\vcr,\tau) = 
e^{i \xi (\vcr,\tau,s)} e^{\lambda_{T,s}(\tau) + \frac{1}{2}\lambda_{L,s}(\tau)} 
\psi^{(0)}_{n,s} (e^{\lambda_{T,s}(\tau)} \vcr_T; e^{\lambda_{L,s}(\tau)} r_3),
\label{scwf}
\end{equation}
with
\begin{equation}
\xi (\vcr,\tau,s) = \frac{\hbar}{m} \left[
 \frac{1}{2} {\dot \lambda}_{T,s}(\tau) \vcr_T^2  
+ \frac{1}{2} {\dot \lambda}_{L,s}(\tau) r_3^2 \right]  
\end{equation}
where $\psi_{n,s}^{(0)}$ is the wave-function in the ground state,
$\lambda_{T,L}(s)$ are the time-dependent collective 
coordinates for the oscillation, and ${\dot \lambda}_{T,L}(s)$ are 
the time-derivatives of $\lambda_{T,L}(s)$ .
The factor $exp(i\xi)$ is the Gallilei transformation factor
which is determined to satisfy the continuum equation.

Let us obtain the total energy under the above scaling by
substituting the wave-function (\ref{scwf})
into the total energy functional (\ref{etot}).
The total energy becomes
\begin{eqnarray}
E_T &=& \frac{\hbar^2}{2 m} 
\sum_{s=1,2} \int d^3 r \left\{ (\nabla \xi)^2 \rho_s + 
e^{2 \lambda_{T,s}} ( \tau^{(0)}_{s,1} + \tau^{(0)}_{s,2} )
+  e^{2 \lambda_{L,s}}  \tau^{(0)}_{s,3} \right\}
\nonumber \\
&&+ \frac{1}{2} m^2 \sum_{s=1,2} 
\int d^3 r \left\{ e^{- 2 \lambda_{T,s} } \Omega_T^2 (r_1^2 + r_2^2)
+ e^{-2 \lambda_{L,s}} \Omega_L^2 r_3^2 )\right\} \rho_s^{(0)} (\vcr) , 
\nonumber \\
&& + g e^{2 \lambda_{T,1} + \lambda_{L,1} + 2\lambda_{T,2}+
 \lambda_{L,2}}
\int d^3 r~ \rho_1^{(0)} (e^{\lambda_{T,1}} \vcr_T; e^{\lambda_{L,1}} r_3)
 \rho^{(0)}_2 (e^{\lambda_{T,2}} \vcr_T; e^{\lambda_{L,2}} r_3)
\end{eqnarray}
where the superscript $(0)$ indicates $\tau_{s,j}$ and $\rho_s$ 
at the ground state.

In this work we use the Thomas-Fermi (TF) approximation to evaluate the above
total energies.
The TF approximation makes the spherical symmetric in the momentum
distribution at each position and gives the following relation:
\begin{equation}
\tau^{(0)}_{s,1} = \tau^{(0)}_{s,2}  = 
\tau^{(0)}_{s,3} = \frac{1}{10 m} (6 \pi^2)^{\frac{2}{3}} {\rho_s^{(0)}}^{\frac{5}{3}} .
\end{equation}
In addition the ground state densities $\rho^{(0)}_s$ become functions
of $\Omega_T (r_1^2 + r_2^2 ) + \Omega_L r_3^2$ .

Here we can then make a change of variables to simplify the
appearance of the Thomas-Fermi equations as follows.
\begin{eqnarray}
x_{j} &=& 
(\frac{m^2 g_0 \Omega_j}{ 3 \pi^2 \hbar^3}) r_{j} ~~~~~ (j=1 \sim 3) , 
\nonumber \\
n_s & = &  \left( \frac{m g_0}{\hbar^2} \right)^3 
\frac{2}{9 \pi^4} \rho_s^{(0)}  ~~~~~ (s=1,2)  ,
\nonumber \\
{\tilde E}_T &=& \frac{4 m^{12} g_0^8 \Omega_L\Omega_T^2}
{(3 \pi^2)^7 \hbar^{21}}  E_T ,
\nonumber \\
g &=& \frac{g_F}{g_0}
\label{etotsc}
\end{eqnarray}
where $g_0$ is a certain coupling constant which is chosen later.

With the scaled variables and the Thomas-Fermi
approximation,
the expression for the total energy at the ground state is 
\begin{equation}
{\tilde E}^{(0)}_T = \int d^3 x ~ \{ \sum_{s=1,2}
( \frac{3}{5} n_s^{\frac{5}{3}} + x^2 n_s ) + g n_1 n_2 \} ,
\label{etotTFsg}
\end{equation}
where $x^2 = |\vx|^2$. 
The TF equations for the densities $n_{1,2}$ are derived by variation of
the energy with a constrain on the fermion numbers.
This yields
\begin{eqnarray}
n_1^{\frac{2}{3}} + g n_2 &=& e_1 - x^2 = e_f + {\tilde B} - x^2 
\nonumber \\
n_2^{\frac{2}{3}} + g n_1 &=& e_2 - x^2 = e_f - {\tilde B} - x^2 
\label{TFeq}
\end{eqnarray}
In this equation $e_1$ and  $e_2$ are the Lagrange multiplier to
constraint the numbers of the fermion-1, $N_1$ and -2, $N_2$, respectively,
and they are rewritten with $e_f$ and ${\tilde B}$;
these $e_f$ and ${\tilde B}$  have the meanings of the 
Fermi energy and the external magnetic field in fermion gas when 
 the two components of fermions are described with 
two different spin states of one fermion.

In this expression the ground state is determined by the three variables,
$g$,  $e_f$ and ${\tilde B}$, but the two variables, $g$ and $e_f$, are
not independent because of the scale symmetry in eq.(\ref{TFeq})
\cite{SY,ToBe}.
In this work we choose $g_0$ as a critical coupling constant 
of the paramagnetic
and ferromagnetic phase transition at a fixed chemical potential in
the symmetric system $\rho_1^{(0)} = \rho_2^{(0)}$.  
With this choice
the fermi energy is fixed to be $e_f = 20/27$. 
The solution of these equations is discussed in detail in ref.~\cite{ToBe}.  
With fixed values of $e_f=20/27$ and ${\tilde B} = 0$, 
the spin phase of the two component fermi gas 
turns to be 'paramagnetic' ($g < 1$) , 'ferromagnetic'($g > 1$) 
in order as the coupling constant, $g$, increases.

Furthermore the total energy in collective oscillations
is given as
\begin{eqnarray}
{\tilde E}_T &=& 
\frac{1}{2} \sum_s {\tilde X}_s \left\{ 
\frac{2 e^{- 2 \lambda_{T,s}}}{3 \omega_T^2}{\dot \lambda}^2_{T,s}
+ \frac{e^{-2 \lambda_{L,s}}}{3 \omega_L^2}{\dot \lambda}^2_{L,s}
\right\}
\nonumber \\
&+& \sum_s \left\{ 
\frac{2 e^{2 \lambda_{T,s}} +  e^{2 \lambda_{L,s}}}{3} {\tilde T}_s
+  \frac{2 e^{-2 \lambda_{T,s}} +  e^{-2\lambda_{L,s}}}{3} {\tilde X}_s
	   \right\}+ \tlV_{ff}
\label{etotTFs}
\end{eqnarray}
with
\begin{eqnarray}
{\tilde T}_s &=&   \frac{3}{5} \int d^3 x ~  n_s^{\frac{5}{3}} (\vx)
\nonumber \\
{\tilde X}_s &=&  \int d^3 x ~ x^2 n_s (\vx)
\end{eqnarray}
where $x^2 = |\vx|^2$, $\omega_T = \Omega_T / \Omega_F$
and  $\omega_L = \Omega_L / \Omega_F$ 
($\Omega_F \equiv \sqrt{\Omega_T^2 \Omega_L}$). 
The interaction energy $\tlV_{ff}$ appears
\begin{eqnarray}
{\tilde V}_{ff} &=& g 
e^{2\lambda_{T,1} + \lambda_{L,1}+ 2\lambda_{T,2} + 2 \lambda_{L,2}} 
 \int d^3 r ~ n_1(e^{\lambda_{T,1}}\vx_T,e^{\lambda_{L,1}} x_3 )
 n_2(e^{\lambda_{T,2}}\vx_T,e^{\lambda_{L,2}} x_3 )
\end{eqnarray}

In order to describe the collective oscillations,
we obtain the variation of the total energy up to the order 
of $O(\lambda^2$) as
\begin{equation}
\Delta E_T \equiv E_T - E^{(0)}_T 
\approx \frac{1}{2} \tdvlamb B \dvlamb
+ \frac{1}{2} \tvlamb C \vlambda
\end{equation}
with
\begin{equation}
B = \left( \begin{array}{cccc} \frac{4 \tlX_1}{3 \omega_T^2} & 0 & 0 & 0 \\
 0 & \frac{2 \tlX_1}{3 \omega_L^2} &  0 & 0 \\
 0 & 0 &  \frac{4 \tlX_2}{3 \omega_T^2} &  0 \\
 0 & 0 &  0 & \frac{2 \tlX_2}{3 \omega_L^2}  \\
 \end{array} \right)
\end{equation}
\begin{eqnarray}
C &=& \left( \begin{array}{cccc} 
\frac{8}{3}(\tlT_1 + \tlX_1)  & 0 & 0 & 0 \\
0 & \frac{4}{3}(\tlT_1 + \tlX_1)  & 0 & 0 \\
0 & 0 & \frac{8}{3}(\tlT_2 + \tlX_2) & 0 \\
0 & 0 & 0 & \frac{4}{3}(\tlT_2 + \tlX_2) \\
 \end{array} \right) 
\nonumber \\
&& ~~~~~~~~~~~ + \left( \begin{array}{cccc} 
~- \frac{4}{3}V_1 - \frac{8}{15}V_3 ~& 
~ - \frac{2}{3}V_1 - \frac{2}{15}V_3~ &
~ \frac{8}{15} V_3 &  \frac{2}{15} V_3 ~ \\

~ - \frac{2}{3} V_1 - \frac{2}{15} V_3 ~ & 
~ - \frac{1}{3} V_1 - \frac{1}{5} V_3 ~ & 
~ \frac{2}{15} V_3 ~ & \frac{1}{5} V_3 ~ \\

~ \frac{8}{15} V_3 ~ & ~\frac{2}{15} V_3 ~ & 
~ -\frac{4}{3} V_2 - \frac{8}{15} V_3 ~ &  
~ -\frac{2}{3} V_2 - \frac{2}{15} V_3 ~\\

~\frac{2}{15} V_3 & \frac{1}{5} V_3 ~ & ~ 
-\frac{2}{3} V_2 - \frac{2}{15} V_3 ~
&  - \frac{1}{3} V_2 - \frac{1}{5} V_3 \\
 \end{array} \right) 
\end{eqnarray}
with
\begin{eqnarray}
V_{1} &=& g \int d^3 x ~  \left( x n_1 \frac{\partial n_2}{\partial x}
			  \right) ,
\\
V_{2} &=& g \int d^3 x ~ \left( x \frac{\partial n_1}{\partial x} n_2
			 \right) , 
\\
V_{3} &=& g \int d^3 x ~ 
 \left( x^2 \frac{\partial n_1}{\partial x} 
\frac{\partial n_2}{\partial x} \right) ,
\end{eqnarray}
where the vector $\vlambda$ is defined as
$\tvlamb = ( \lambda_{T,1}, \lambda_{L,1}, \lambda_{T,2}, \lambda_{L,2} )$.
The terms propritional to $\vlambda$ disappears because of the virial theorem. 

Then the classical equation of motion for $\vlambda$ is harmonic, 
giving rise to the following eigenvalue equation for the oscillation
frequencies,
\begin{equation}
\left[ B \omega^2 - C  \right] \vlambda  = 0 .
\label{eigEq}
\end{equation}

Before performing actual calculations, we should discuss the collective
oscillations in some extreme conditions. 

In the largely deformed system, $\omega_L >>1$ or $\omega_T >>1$,
first, the four oscillation modes are decoupled into two groups,
two oscillations modes in the longitudinal directions, and the other two modes 
in the transverse directions.
We can obtain these decoupled modes by taking 
$C_{12} = C_{14} = C_{23}=C_{34}=0$.
In the symmetric system, $n_1 = n_2$, furthermore, 
the two kinds of modes are further decoupled into the in-phase 
($\lambda_{L(T),1} = \lambda_{L(T),2}$)
and the out-of-phase 
($\lambda_{L(T),1} = -\lambda_{L(T),2}$) modes;
namely the four excited oscillations are decoupled into 
the in-phase transverse breathing (ITB), 
in-phase longitudinal breathing (ILB), out-of-phase transverse breathing
(OTB) and out-of-phase longitudinal breathing (OLB) modes.

The oscillation frequencies in the symmetric system become
\begin{eqnarray}
{\omega_T^{in}}^2 = \frac{2(\tlT_1 + \tlX_1) -  V_1 }{\tlX_1} \omega_T^2
&,~~&
{\omega_T^{out}}^2 = \frac{2(\tlT_1 + \tlX_1) - 
 V_{1} - \frac{4}{5} V_{3} }{\tlX_1} \omega_T^2 ,
\\
{\omega_L^{in}}^2 = \frac{2(\tlT_1 + \tlX_1)
- \frac{1}{2} V_1}{\tlX_1} \omega_L^2
&,~~&
{\omega_L^{out}}^2 = \frac{2(\tlT_1 + \tlX_1)- \frac{1}{2} V_1 
- \frac{3}{5} V_{3} }{\tlX_1} \omega_L^2 ,
\end{eqnarray}
where the superscripts '$in$' and '$out$' indicate
the in-phase and out-of-phase modes, respectively, and
the subscripts 'T' and 'L' denote the oscillations in the transverse and
longitudinal directions, respectively. 
Note that $\tlT_1 = \tlT_2$, $\tlX_1 = \tlX_2$ and so on.

Next we discuss the oscillations in the near-spherical symmetric system, 
$\omega_L \approx \omega_T \approx 1$.
In this system the oscillation modes are decoupled into the monopole 
and quadrupole oscillation modes.

By taking the collective coordinates to be 
$\lambda_{L,i} = \lambda_{T,1}=\lambda_i$ , 
we can obtain the frequencies of  the monopole oscillations with 
the following equation:
\begin{equation} 
\left[ \left( \begin{array}{cc} 
2\tlX_1 & 0  \\
0 &  2 \tlX_2   \end{array} \right) \frac{\omega_2}{\Omega_M^2}
- \left( \begin{array}{cc} 
4(\tlT_1 + \tlX_1) - 3V_1 - V_3 & V_3  \\
V_3  & 4(\tlT_2 + \tlX_2)  -3 V_2 - V_3
 \end{array} \right)  \right] 
\left( \begin{array}{c} \lambda_1 \\ \lambda_2 \end{array} \right)   = 0 
\end{equation}
with
\begin{equation}
\frac{1}{\Omega_M^2} = \frac{1}{3}(\frac{2}{\omega_T^2} + \frac{1}{\omega_L^2}).
\end{equation}

By taking the collective coordinates to be 
$\lambda_{T,i} = - \lambda_{L,i}/2 = \lambda_i$, similarly,
we can also calculate the frequencies of  the quadrupole oscillations with
\begin{equation}
\left[ 
\left( \begin{array}{cc} 
\tlX_1 & 0  \\ 0 &  \tlX_2  
 \end{array} \right)\frac{\omega^2}{\Omega^2_Q}
- \left( \begin{array}{cc} 
2 (\tlT_1 + \tlX_1) - \frac{1}{5} V_3 ~ & \frac{1}{5} V_3 ~ \\
\frac{1}{5} V_3 ~ & 2 (\tlT_2 + \tlX_2)  - \frac{1}{5} V_3
 \end{array} \right) \right] 
\left( \begin{array}{c} \lambda_1 \\ \lambda_2 \end{array} \right)  = 0 
\end{equation}
with
\begin{equation}
\frac{1}{\Omega_Q^2} = 
\frac{1}{3}(\frac{1}{\omega_T^2} + \frac{2}{\omega_L^2}) .
\end{equation}

In the symmetric system $\rho_1^{(0)} = \rho_2^{(0)}$, these modes
are also decoupled to the in-phase and the out-of-phase modes.
Then the oscillation frequencies in the symmetric system become
\begin{eqnarray}
{\omega_M^{in}}^2 = \frac{2(\tlT_1 + \tlX_1) - \frac{3}{2} V_{1}
 }{\tlX_1} \Omega_M^2 ,
&~~~&
{\omega_M^{out}}^2 = \frac{2(\tlT_1 + \tlX_1) - \frac{3}{2} V_{1} -
V_{3} }{\tlX_1} \Omega_M^2 .
\\
{\omega_Q^{in}}^2 = \frac{2(\tlT_1 + \tlX_1)}{\tlX_1} \Omega_Q^2,
&~~~&
{\omega_Q^{out}}^2 =  \frac{2(\tlT_1 + \tlX_1) - \frac{2}{5} V_{3} }{\tlX_1}
\Omega^2 .
\end{eqnarray}
where the subscript 'M' and 'Q' indicate the monopole and quadrupole 
frequencies, respectively.

\section{Results}

In this section we discuss the collective oscillations coupled with 
the transverse and longitudinal breathing modes.
As mentioned in the previous section 
we fix the fermi energy to be $e_f = 20/27$, 
which makes the parametric and ferromagnetic spin phase transition at
$g=1$ when ${\tilde B} = 0$.
 
\begin{figure}[ht]
\vspace{0.5cm}
\hspace*{0cm}
\includegraphics[scale=0.8]{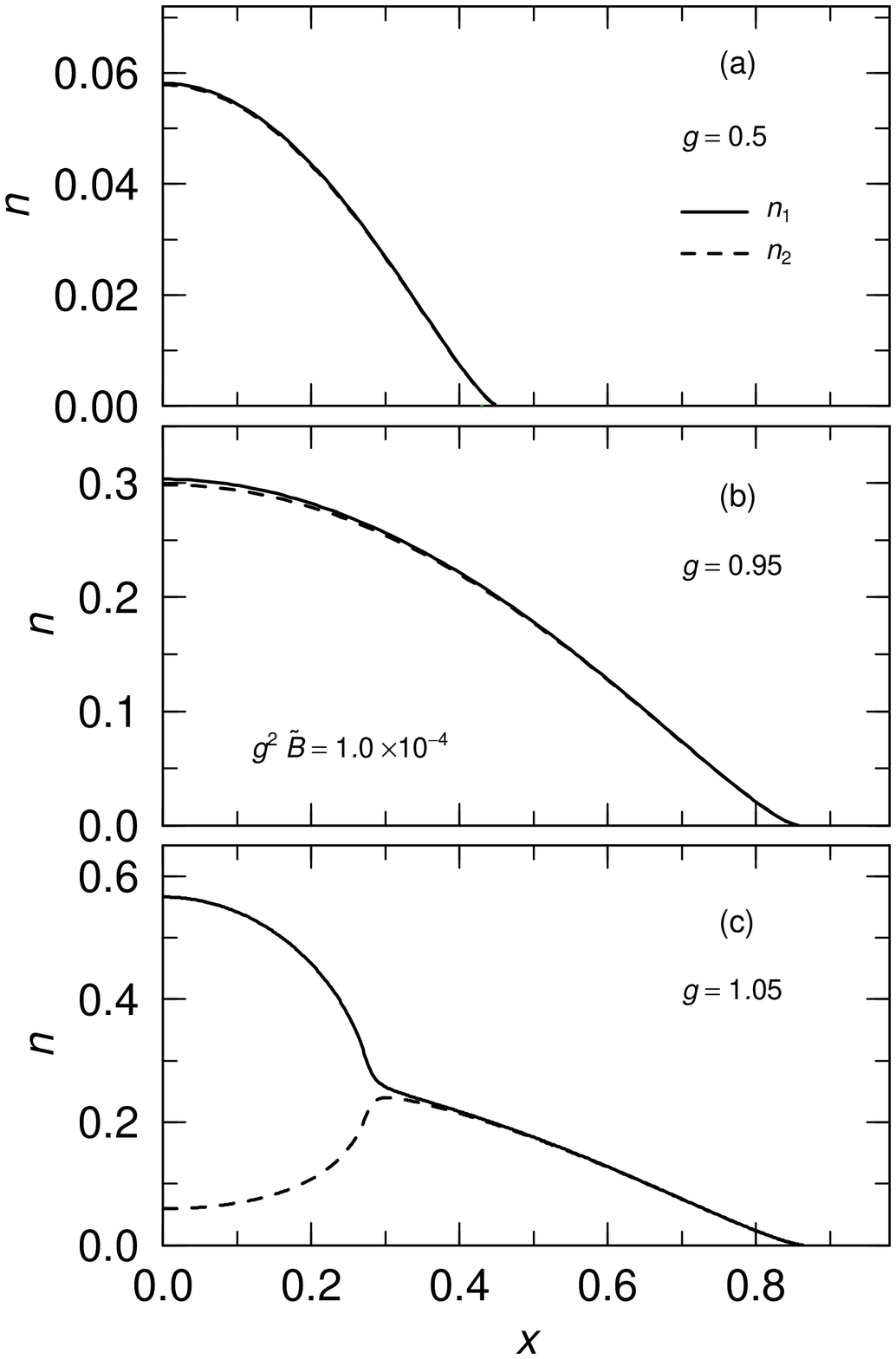}
\caption
{\small 
The scaled density distribution of two components fermion
at $g = 0.5$ (a), $g = 0.95$ (b) and $g = 1.05$ (c)
with with $e_f=20/27$ and $g^2 {\tilde B} = 1.0 \times 10^{-4}$. 
The solid and dashed lines represents the distribution
of the major and minor components, respectively. 
}
\label{frho}
\end{figure}

Density profiles illustrating these three regimes are shown in Fig.
\ref{frho}.  
The panels show the densities as a function of distance $x$
in a weak magnetic field ($g^2 \Btl = 1.0 \times 10^{-4}$)
and at three different interaction strengths:
$g = 0.5$(a), 0.95(b) and 1.05(c).
The solid and dashed
lines represent the scaled density distribution of major and minor components
of fermions.
From top to bottom, the panels show paramagnetic (a,b) and ferromagnetic (c).

Next we calculate the collective oscillations by solving 
eigenvalue equation (\ref{eigEq}).
In this equation we obtain four eigenvalues corresponding to collective
oscillation frequencies.
For convenience we define these four frequencies 
as $\omega_1$,  $\omega_2$,  $\omega_3$ and  $\omega_4$
in order of the frequencies from the larger one to the lower one.
In addition we refer to the four collective modes with the frequencies
 $\omega_1$,  $\omega_2$,  $\omega_3$ and  $\omega_4$ as
modes-1, -2, -3 and -4, respectively.

\begin{figure}[ht]
\vspace{0.5cm}
\hspace*{0cm}
\includegraphics[scale=0.8]{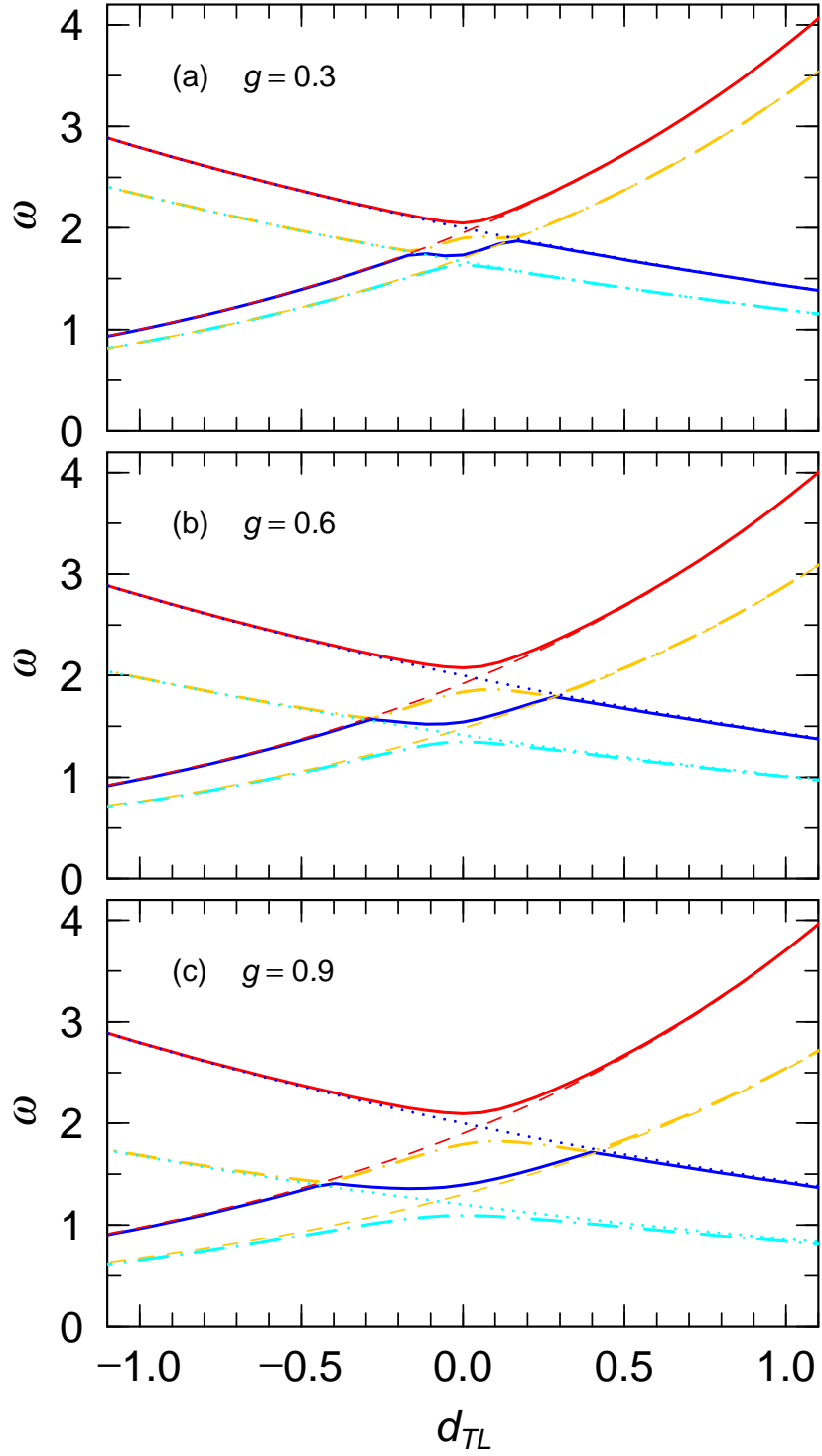}
\caption
{\small 
The frequencies of the collective oscillations
versus $\omega_L/\omega_T$
at $g = 0.3$ (a), $g = 0.6$ (b) and $g = 0.9$ (c)
with $e_f=20/27$ and $g^2 {\tilde B} = 1.0 \times 10^{-4}$. 
The solid lines represent the results with the full calculations,
and the dashed and dotted lines indicate
the frequencies of the transverse and longitudinal breathing modes,
respectively.
}
\label{mdLT1}
\end{figure}

In Fig.~\ref{mdLT1} we show the frequencies of oscillations 
as functions of the deformation parameter $d_{TL} = \ln (\omega_L/\omega_T)$ 
at the coupling constant $g=0.3$ (a), $= 0.6$ (b) and $=0.9$ (c). 
The solid lines represent the first and third frequencies 
$\omega_1$ and $\omega_3$, and the chain-dotted lines indicate
$\omega_2$ and $\omega_4$.
In these calculation we take the magnetic field to be  
$g^2 {\tilde B} = 1.0 \times 10^{-4}$, where the system is almost
symmetric between the two component, $N_1 \approx N_2$.
In the same figures, furthermore, we also plot the frequencies 
of the transverse and longitudinal breathing modes 
with the dashed and dotted lines, respectively.

First we see the strong level mixing between modes-2 and -3 at
two deformation parameters, $d_{TL} = d_1$ and $d_2$ ($d_1 < 0 < d_2$). 
When $d_{TL} <d_1$ and $d_2 < d_{TL}$, the frequencies in the full calculation
(solid lines) are almost the same as those in the calculations decoupled with
the transverse and longitudinal modes (dashed and dotted lines), 
though they are quite different when $d_1 < d_{TL} < d_2$.

\begin{figure}[ht]
\vspace{0.5cm}
\hspace*{0cm}
\includegraphics[scale=0.8]{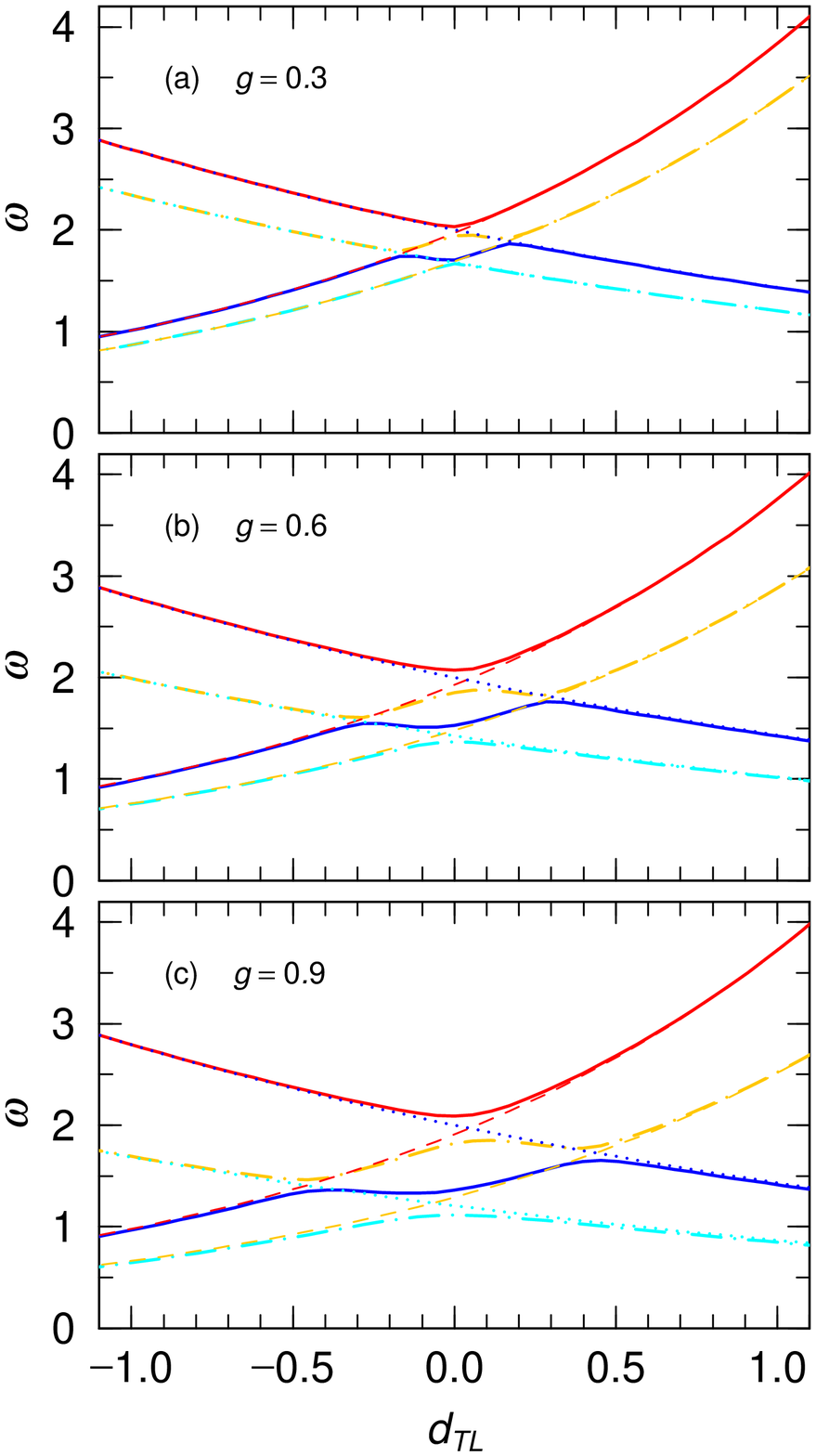}
\caption
{\small 
Same as Fig.~\ref{mdLT1}, but with
$g^2 {\tilde B} = 1.0 \times 10^{-2}$. 
}
\label{mdLT2}
\end{figure}

In Fig.~\ref{mdLT2} we also give the same quantities but with
$g^2 {\tilde B} = 1.0 \times 10^{-2}$.
The system is not symmetric $n_1 \neq n_2$, but the results are quite
similar to those in Fig.~\ref{mdLT1} though the difference in the frequency
at the level mixing points between the modes-2 and -3 is a little
larger than those in Fig.~\ref{mdLT1}. 

For future discussions we refer to the three deformation parameter regions,
$d_{TL} < d_1$, $d_1< d_{TL} < d_2$ and  $d_2 < d_{TL}$ as
the prolate deformed, the semi-spherical and 
the prolate deformed regions, respectively.
The transverse and longitudinal breathing oscillation modes are almost
decoupled in both the oblate and prolate deformed regions,  
while these modes are coupled in the spherical region.

\begin{figure}[ht]
\vspace{0.5cm}
\hspace*{0cm}
\includegraphics[scale=0.8]{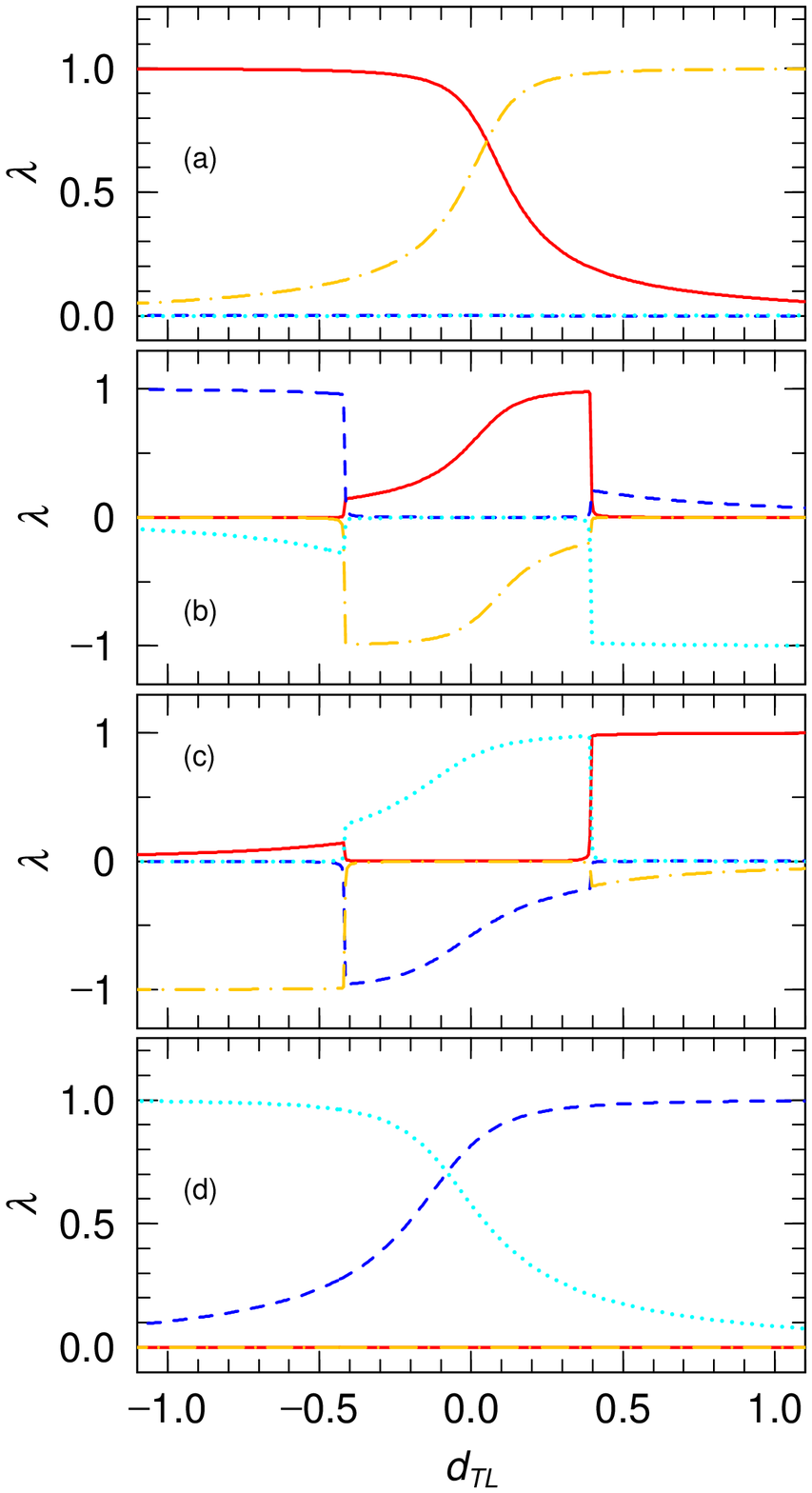}
\caption
{\small 
Components of the eigenvectors
of mode-1 (a), mode-2 (b), mode-3 (c) and mode-4 (d)
at $g= 0.9$ with $g^2 {\tilde B} = 1.0 \times 10^{-4}$. 
The solid and chain-dotted lines represent the
in-phase transverse and longitudinal components, respectively,
and the dashed and dotted lines denote the out-of-phase 
transverse and longitudinal components, respectively. 
}
\label{evc1}
\end{figure}

In order to know the level mixing behaviors more,  
we calculate the four following components:
\begin{eqnarray}
\lambda_T^{in} = \frac{1}{\sqrt{2}} ( \lambda_{T,1} + \lambda_{T,2} ) ,
&~~~&
\lambda_L^{in} = \frac{1}{\sqrt{2}}
( \lambda_{L,1} + \lambda_{L,2} ) ,\\
\lambda_T^{out} = \frac{1}{\sqrt{2}}
( \lambda_{T,1} - \lambda_{T,2} ) ,
&~~~&
\lambda_L^{out} = \frac{1}{\sqrt{2}}
( \lambda_{L,1} - \lambda_{L,2} ) ,
\end{eqnarray}
where $\lambda_T^{in}$, $\lambda_L^{in}$, $\lambda_T^{out}$ and  
$\lambda_L^{out}$
are components of the ITB, ILB, OTB and OLB modes, respectively.
In Fig.~\ref{evc1} we show the components of the eigenvectors
at the mode-1 (a), the mode-2 (b), the mode-3 (c) and the mode-4 (d)
with $g = 0.9$ and $g^2 {\tilde B} = 1.0 \times 10^{-4}$.
The solid, chain-dotted, dashed and dotted lines represent
 $\lambda_T^{in}$, $\lambda_L^{in}$, $\lambda_T^{out}$ and  
$\lambda_L^{out}$, respectively.

The results show that the phase  of the modes-1 and -4 
are in-phase in all $d_{TL}$ regime.
When the system is oblate deformed ($d_{TL} << 0$).
the mode-1  is almost the ITB mode.
As $d_{TL}$ increase, 
the system approaches the spherical, the ILB component in the mode-1 mode 
gradually becomes larger. 
When the system becomes spherical ($d_{TL} = 0$),
the mode-1 becomes a mixed mode with the ITB and ILB in the same ratio:  
namely the mode-1 is the monopole oscillation. 
As the system is further deformed to the prolate shape,
the transverse component decreases and the longitudinal one 
increases and the oscillation becomes the in-phase longitudinal
oscillation in  $\omega_L /\omega_T \rightarrow \infty$ limit.

The mode-4 has the same behavior, but its phase is out-of-phase;
the OTB mode when $d_{TL} < d_1$,  
the out-of-phase monopole mode at $d_{TL} =0$
and the OLB mode when $d_{TL} > d_2$. 

The variations of the modes-2 and -3 show more complicated behaviors.
The mode-2 is mainly the OTB mode in the oblate deformed region 
($d_{TL} < d_1$), 
and the OLB mode in the prolate deformation region ( $d_{TL} > d_2$).
The mode-3 is the ILB mode in the oblate deformation, and
the ITB mode in the prolate deformation.
At $d_{TL} = d_1$ and $d_2$, the modes-2 and -3 make the level mixing
and exchange their roles.
Namely the modes-2 and -3 are in-phase out-of-phase oscillations
in the semi-spherical region, respectively.
As $d_{TL}$ approaches zero, the TB and LB oscillation modes are mixed.
At $d_{TL} = 0$, the mode-2 becomes the in-phase quadrupole oscillations,
($\lambda_L^{in} = - 2 \lambda_T^{in}$)
and the mode-3 becomes the out-of-phase ones, respectively.

Thus we know that
the four modes agree with the frequencies of the TB and LB modes,
when $\omega_L/\omega_T >>1 $ ($d_{TL} >> 0$),
and  $\omega_T/\omega_L >>1 $ ($d_{TL} << 0$).
On the other hand these modes show different behaviors in 
the nearly spherical symmetric system 
($d_{TL} \approx 0$),
where the collective oscillations are described as the coupled modes
of the monopole and quadrupole oscillations.

\begin{figure}[ht]
\vspace{0.5cm}
\hspace*{0cm}
\includegraphics[scale=0.8]{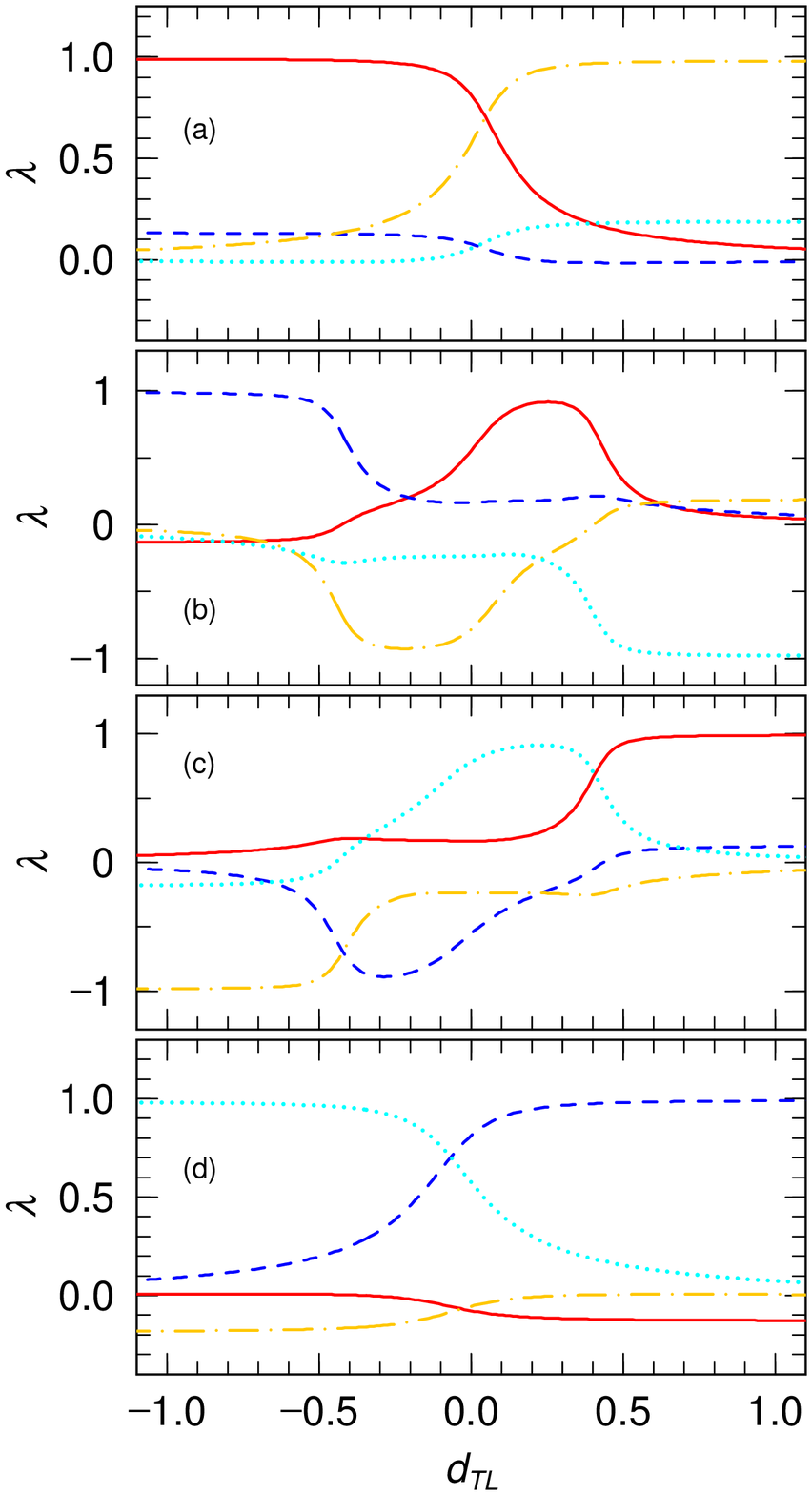}
\caption
{\small 
Same as Fig.~\ref{evc1}, but with
$g^2 {\tilde B} = 1.0 \times 10^{-2}$. 
}
\label{evc2}
\end{figure}

In Fig.~\ref{evc2}, furthermore, we show the same quantities in
Fig.~\ref{evc1} but with  $g^2 {\tilde B} = 1.0 \times 10^{-2}$.
The feature discussed above is less clear, but also seen
in the case of the strong magnetic field.

When $g^2 {\tilde B} = 1.0 \times 10^{-2}$,
the number asymmetry is not so small, $N_1  \gtrsim N_2$.
The mixing between the in-phase and out-of-phase modes 
is very small in modes-1 and -4.
The similar result was shown for the monopole oscillation in the
spherical trap.
Even in modes-2 and -3, this mixing is seen only in the
region near $d_{TL} \approx d_1$ and  $d_{TL} \approx d_2$.
Hence we can disregard the mixing between the in-phase and out-of-phase
modes in qualitative discussions,
and consider that modes-2 and -3 exchange their roles around  
 $d_{TL} \approx d_1$ and  $d_{TL} \approx d_2$.

\begin{figure}[ht]
\vspace{0.5cm}
\hspace*{0cm}
\includegraphics[scale=0.7, angle=270]{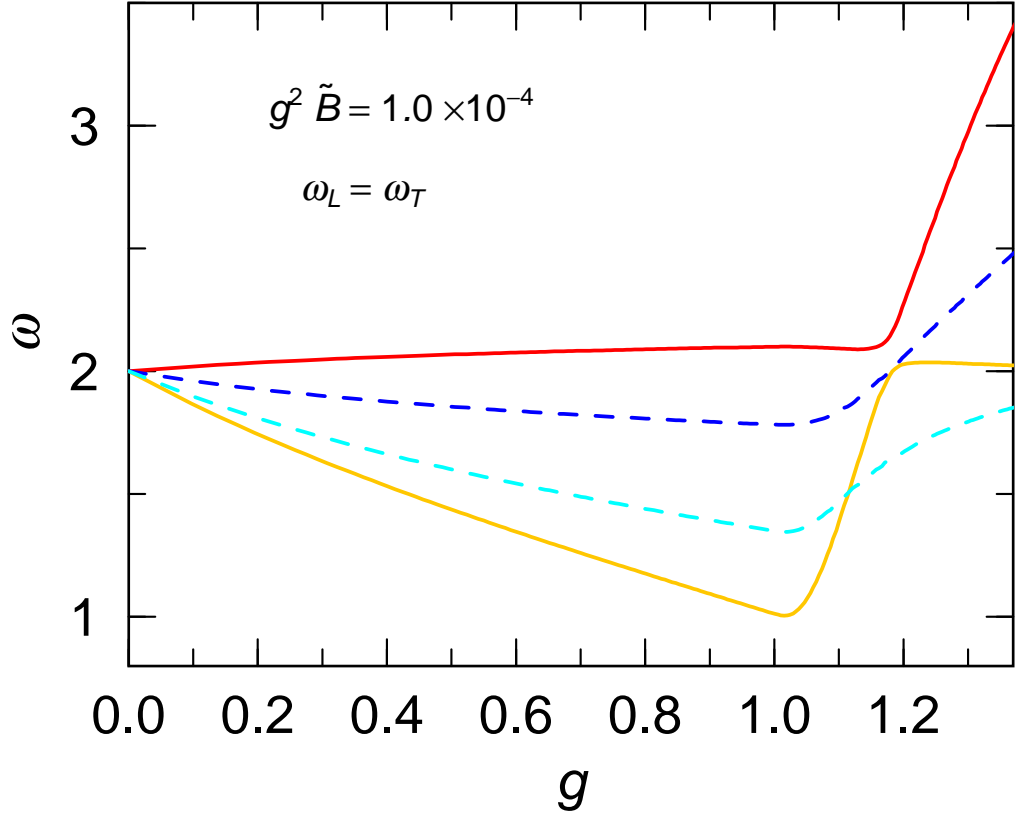}
\caption
{\small 
The frequency of the monopole and quadrupole oscillations
with the external magnetic fields with 
$g^2 {\tilde B} =$ $1.0 \times 10^{-4}$.
The solid and dashed lines represent the results of
the monopole and quadrupoles  modes, respectively. }
\label{spMQ}
\end{figure}

In order to know more details of the deformation dependence, we should
examine relations between the frequencies 
of the collective oscillations and the coupling constant $g$. 
First we show the frequencies of the monopole 
and quadrupole oscillations versus the coupling constant $g$
in the spherical symmetric system ($\omega_L = \omega_T$) with the 
$g^2 {\tilde B} = 1.0 \times 10^{-4}$ in Fig.~\ref{spMQ}. 
The solid and dashed lines represent the results 
of the monopole and quadrupole oscillations.
In the completely spherical symmetric system the monopole and quadrupole
states are decoupled, and there is no level mixing between these two modes. 
In the paramagnetic phase ($g \lesssim 1.0$), larger frequencies correspond
to the in-phase oscillation modes both in the monopole and quadrupole modes.
The monopole modes make the level mixing  between the in-phase
and out-of-phase modes in the ferromagnetic phase  ($g \lesssim 1.0$), 
while the quadrupole modes do not show the level mixing.
As the coupling becomes larger,
the frequency of the oscillation modes except the in-phase monopole
oscillation mode decreases monotonically until $g \approx 1.0$, and  
increases above that, while the frequencies of the in-phase 
monopole oscillation are almost invalid.
When the coupling constant increases further,
the level mixing between the two modes of the monopole oscillation
occurs.

\begin{figure}[ht]
\vspace{0.5cm}
\hspace*{0cm}
\includegraphics[scale=0.7, angle=270]{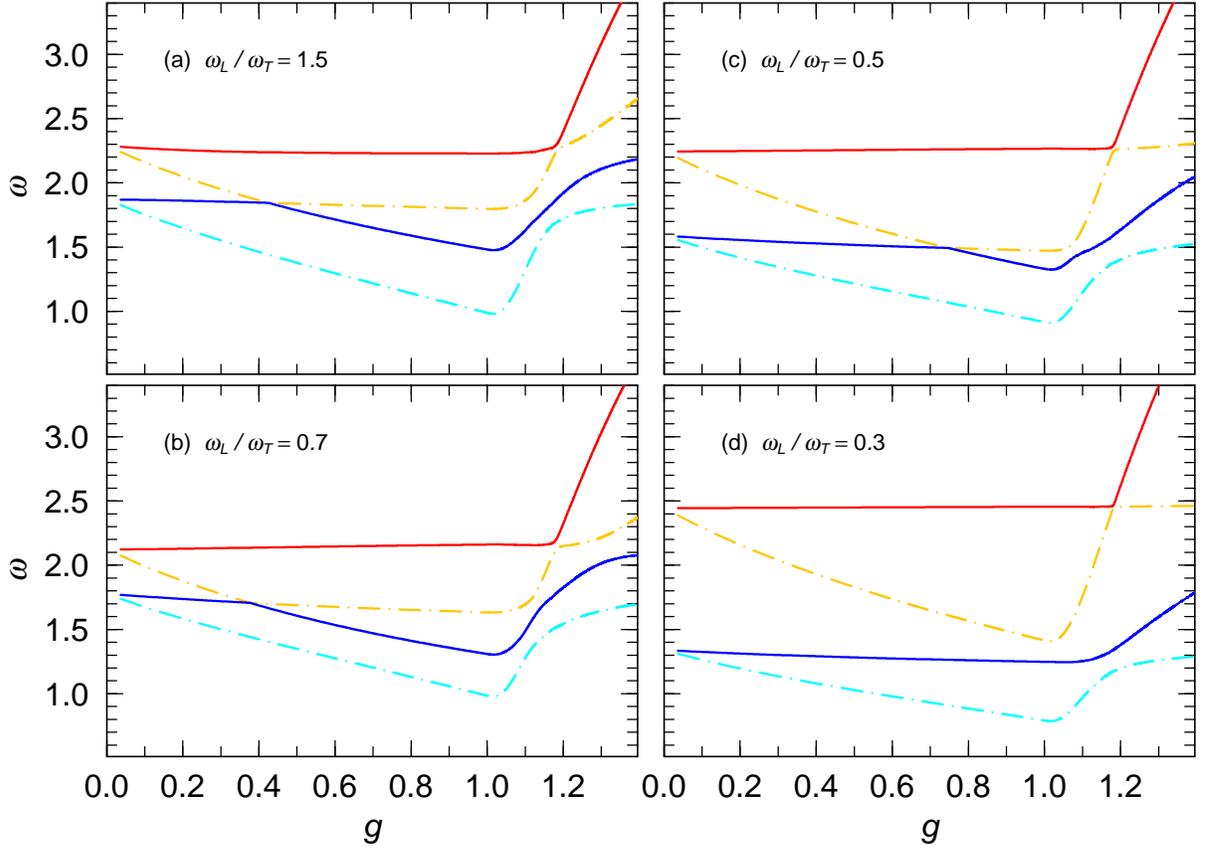}
\caption
{\small 
The frequencies of the collective oscillations
versus the coupling constant $g$ with the external magnetic fields with 
$g^2 {\tilde B} =$ $1.0 \times 10^{-4}$
for $\omega_L/\omega_T = 1.5$ (a),  0.7 (b),  0.5 (c) and 0.3 (b).
The solid lines represent the results of the mode-1 and the mode-3,
and the chain-dotted lines indicate those  of the mode-2 and the mode-4.}
\label{spclMQ}
\end{figure}

Next we study the non-symmetric system, where all the four modes can
be mixed.
In Fig.~\ref{spclMQ}
we show the frequencies of the mode-1, -2, -3 and -4
as the functions of the coupling constant with 
$\omega_L/\omega_T = 1.5$ ($d_{TL} = 0.41$) (a), 
$\omega_L/\omega_T =0.7$ ($d_{TL}=-0.36$) (b), 
$\omega_L/\omega_T =0.5$ ($d_{TL}=-0.69$) (c)
 and $\omega_L/\omega_T =0.3$ ($d_{TL}=-1.2$) (d).
In these figures we see again that the levels between modes-2 and 
-3 are mixed at a certain $g$.
When $\omega_L/\omega_T=0.7$ (Fig.~\ref{spclMQ}), for example, 
the frequencies of modes-1 and -2 are $\omega = 2\omega_T$,
and those of modes-3 and -4 are $\omega = 2\omega_L$ at $g=0$.
As the coupling becomes larger, the frequencies of modes-2 and -4 
monotonously decreases while those of modes-1 and -3 is almost 
invalid, and then the the level mixing between modes-2 and -3 
occurs at $g \approx 0.37$.  

As the deformation parameter $d_{TL}$ increases, 
the difference between $\omega_T$ and
$\omega_L$ becomes larger, and the level mixing occurs at larger
coupling constant.
As the deformation parameter further increases,  
the level mixing disappears when $d_{TL} < d_1$ 
because the minimum value of $\omega_2$ is larger than $\omega_3$,
and the level mixing cannot occur.
The $\omega_2$ becomes minimum at the border between the paramagnetic and
ferromagnetic phases $g \approx 1$, where
the level mixing occurs when $d_{TL} = d_1$.

\begin{figure}[ht]
\vspace{0.5cm}
\hspace*{0cm}
\includegraphics[scale=0.7,angle=270]{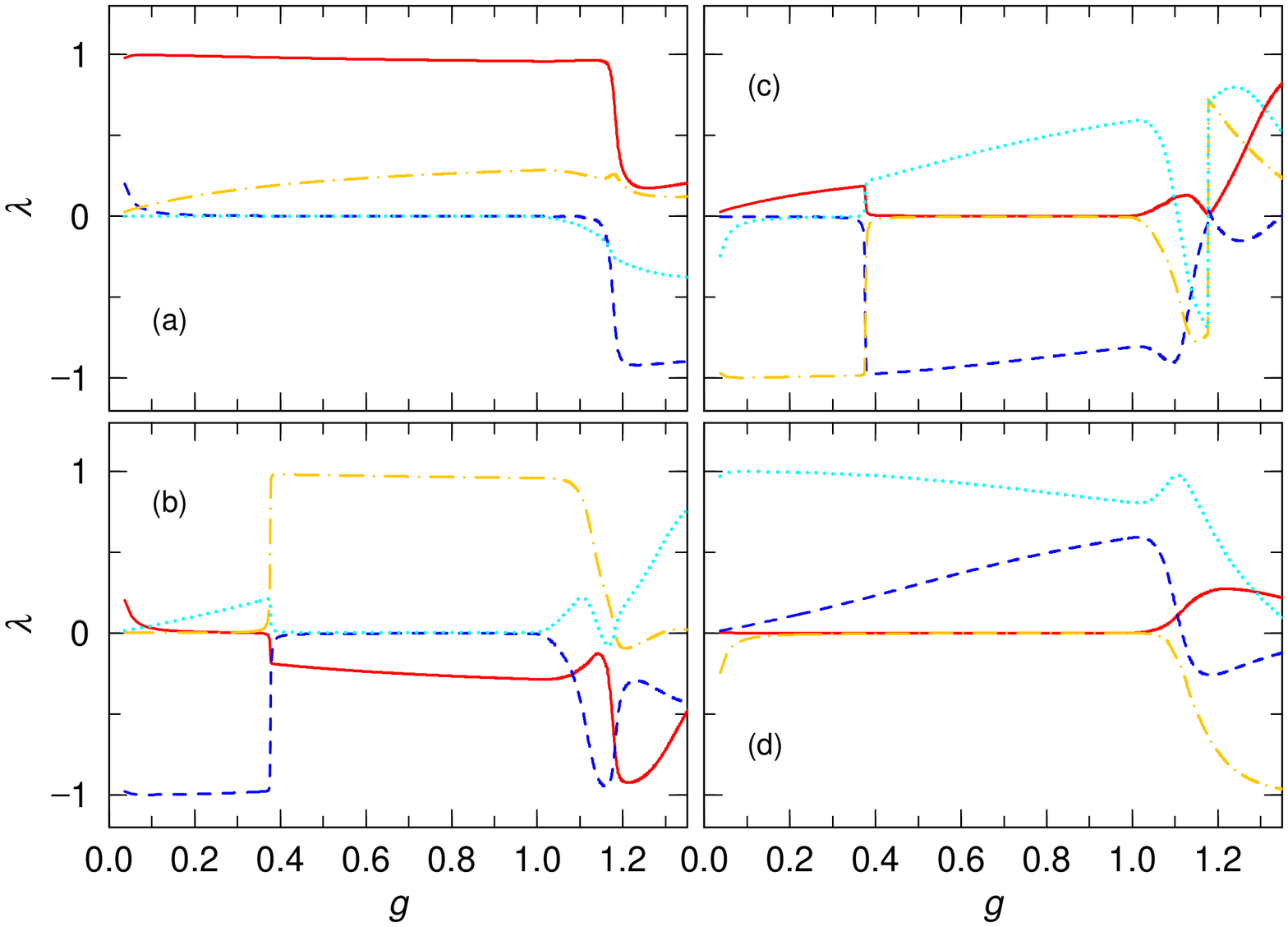}
\caption
{\small 
Components of the eigenvectors
of mode-1 (a), mode-2 (b), mode-3 (c) and mode-4 (d)
with $g^2 {\tilde B} = 1.0 \times 10^{-4}$ and  $\omega_L/\omega_T = 0.7$. 
The solid and chain-dotted lines represent the
in-phase transverse and longitudinal components, respectively,
and the dashed and dotted lines denote the out-of-phase 
transverse and longitudinal components, respectively. }
\label{spvcG}
\end{figure}

In Fig.~\ref{spvcG}  we show the components of the eigenvectors
at the mode-1 (a), the mode-2 (b), the mode-3 (c) and the mode-4 (d)
with $g^2 {\tilde B} = 1.0 \times 10^{-4}$ and $\omega_L / \omega_T = 0.7$.
The mode-1 and the mode-4 are almost ITB and OLB modes
in the weak coupling region, respectively.
As the coupling, $g$, increases, the ILB and OTB components are
gradually mixed in both the modes.
In the case of the mode-1, this mixing is small, while in the mode-4 
the OLB and OTB components are more largely mixed, particularly 
around $g \approx 1$. 

In this figure, furthermore, we can also see the level mixing between
modes-2 and -3.
At $g=0$ the mode-2 and the mode-3 are the ILB and OTB modes.
As the coupling increases, the ITB and OLB components are more
mixed in the mode-2 and the mode-3, respectively.
At $g \approx 0.37$ the level mixing occurs.
The strength of the components is exchanged between the modes-2 and -3.

\section{Summary}

In this paper we study the collective breathing oscillations in the
deformed traps.
The oscillations contain the four kinds of the oscillation, 
ITB, OTB, ILB and OLB modes.
When the shape of the traps are largely deformed, either oblate or prolate,
these modes are decoupled.
As this shape becomes close to being spherical,
the transverse and longitudinal modes are coupled,
and make the monopole and quadrupole oscillations.
The borders between these coupling and decoupling are at 
$d_{TL} \approx  d_1$  and $d_{TL} \approx  d_2$,
where   the OTB and ILB modes make level crossing. 

In the deformed traps the frequencies of the ITB and ILB modes agree
with twice of the trap frequencies and are not so different from those of
the non-interacting system.
The frequencies of the OTB and OLB modes show some coupling dependence
and give information of the interaction,
Their qualitative behaviors are however similar to those in the dipole
oscillation \cite{ToBe}.

In the semi-spherical traps, on the other hand, the collective oscillations
exhibit more complicated behavior because of the level mixing.
Hence we can get significant information on many body system of fermi
gas by investigate breathing oscillations with the various fermion-fermion
couplings and deformations of the trap near being spherical 
($0.4 \lesssim \omega_L/\omega_T \lesssim 1.6$).

In this work we do not discuss extremely deformed system 
($\omega_T \rightarrow \infty$ or $\omega_L \rightarrow \infty$).
In such extreme conditions the system becomes a quasi-low dimension one 
\cite{Petrov}, 
and the collective oscillation can be expected to 
show quite different behaviors from those in our work \cite{Nishimura1}.

\newpage

\newpage

\newpage

\newpage
\newpage

\newpage

\newpage

\end{document}